\documentclass[twocolumn,showpacs,nofootinbib]{revtex4}

\begin{document}

%\preprint{}

\title{Towards a general solution of the Hamiltonian constraints of General Relativity}

% Explanatory text should go in the []'s, actual e-mail

\author{A. Tiemblo}
\email[]{tiemblo@imaff.cfmac.csic.es}
%\thanks{}
%\affiliation{}

\author{R. Tresguerres}
\email[]{romualdo@imaff.cfmac.csic.es}
%\thanks{}
\affiliation{Instituto de Matem\'aticas y F\'isica Fundamental\\
Consejo Superior de Investigaciones Cient\'ificas\\ Serrano 113
bis, 28006 Madrid, SPAIN}

\date{\today}

\begin{abstract}
The present work has a double aim. On the one hand we call
attention on the relationship existing between the Ashtekar
formalism and other gauge-theoretical approaches to gravity, in
particular the Poincar\'e Gauge Theory. On the other hand we study
two kinds of solutions for the constraints of General Relativity,
consisting of two mutually independent parts, namely a general
three-metric-dependent contribution to the extrinsic curvature
$K_{ab}$ in terms of the Cotton-York tensor, and besides it
further metric independent contributions, which we analyze in
particular in the presence of isotropic three-metrics.
\end{abstract}

%\keywords{}
\maketitle

\section{Introduction}

In what follows we are mainly concerned with the search for a
solution of the Hamiltonian constraints of General Relativity
(GR). However, incidentally we are also interested in showing the
underlying linkage existing between the Ashtekar formalism and the
Hamiltonian Poincar\'e Gauge Theory (PGT) discussed by us in
previous works \cite{Lopez-Pinto:1997aw} \cite{rrdph}. In our
view, the internal $SO(3)$ symmetry affecting Ashtekar's variables
\cite{Ashtekar:1988sw} reveals to be the explicit manifestation of
a wider gauge group. In order to corroborate this fact, we present
the constraints of GR in different formulations, showing the bond
between the PGT, the Ashtekar and the ADM ones
\cite{DeWitt:1967yk}.

After revising previous attempts to solve the constraints of GR,
we derive our main result, consisting of equation
(\ref{solution1}), which provides a solution of both the Gauss law
and the ADM momentum constraint (or Ashtekar's vector constraint)
by expressing the exterior curvature in terms of the three-metric
and its derivatives. Any other solution is defined up to this
metric dependent contribution. We finalize discussing additional
acceptable solutions irreducible to (\ref{solution1}).

As a necessary reference for our discourse, let us first of all
review the basics of the Ashtekar Hamiltonian formulation of
gravity generalized by Barbero \cite{Barbero:1995ap}
\cite{Immirzi:1996dr} \cite{Samuel:2000ue}, where the fundamental
dynamical variables are given by the canonically conjugate pair
$\left(\,E_I{}^a\,,\,A_a^I\,\right)$. (In our notation, small
latin letters $a$, $b$... of the beginning of the alphabet are
assigned to the coordinates of the three-dimensional spatial
slices resulting from a suitable foliation of spacetime, while
capital letters $I$, $J$... of the middle of the alphabet display
internal $SO(3)$ indices running from $1$ to $3$.) The first pair
element consists of the densitized triads
\begin{equation}
E_I{}^a :=e\,e_I{}^a\,,\label{triad}
\end{equation}
defined from the triad components $e_I{}^a\,$ on the spatial
slices, with the corresponding determinant
$e:={1\over{3!}}\,\epsilon ^{abc}\epsilon
_{_{IJK}}\,e_a{}^I\,e_b{}^J\,e_c{}^K$. The second pair element,
that is the momentum conjugate to $E_I{}^a$, is the $SO(3)$
connection
\begin{equation}
A_a^I :=\Gamma _a^I (e) +\beta\,K_a{}^I\,,\label{Ashtekbeta3}
\end{equation}
constituted by two contributions, namely the triad compatible spin
coefficients $\Gamma _a^I (e)$, whose explicit form is given in
(\ref{SO3conn}) with (\ref{SO3connbis}) below, and the extrinsic
curvature tensor $K_a^I$ of the slices (with an index converted
into a triad index), being both combined by means of the
Barbero-Immirzi parameter $\beta$. Explicitly, the spin part of
(\ref{Ashtekbeta3}) is built from the triads $e_I{}^a\,$ as
\begin{equation}
\Gamma _a^I (e):= {1\over 2}\,\epsilon ^{IJK}\hat{\Gamma}_{aJK}
(e)\,,\label{SO3conn}
\end{equation}
with
\begin{equation}
\hat{\Gamma}_{aJK} (e):=-e_{[J}{}^b\,\Bigl(\,\partial _a
e_{bK]}-\Gamma _{ab}{}^c e_{cK]}\,\Bigr)\,,\label{SO3connbis}
\end{equation}
see (\ref{Christoffel}), where $\Gamma _{ab}{}^c$ stands for the
holonomic spatial Christoffel symbol
\begin{equation}
\Gamma _{ab}{}^c :={1\over 2}\,q^{\,cd}\Bigl(\,\partial _a
q_{\,db}+
\partial _b q_{\,da}-\partial _d q_{\,ab}\,\Bigr)\,,\label{Christoffel3}
\end{equation}
expressed in terms of the holonomic three-metric $q_{\,ab}:=\delta
_{IJ}\,e_a{}^I e_b{}^J\,$ defined from the triads.

The constraints of General Relativity in terms of (\ref{triad})
and (\ref{Ashtekbeta3}), as derived in the literature
\cite{Barbero:1995ap} \cite{Immirzi:1996dr} \cite{Samuel:2000ue},
are the weakly vanishing expressions
\begin{eqnarray}
{\cal G}_I &&\,:=\nabla _a{} E_I{}^a\approx\,0\,,\label{Asht1}\\
{\cal V}_a &&\,:= E_I{}^b F^I_{ab}\approx\,0\,,\label{Asht2}\\
\nonumber {\cal S} &&\,:=\,\epsilon _I{}^{JK} E_J{}^a E_K{}^b F^I_{ab}\\
\nonumber -{2(1+\beta ^2 )\over{\beta ^2}}&&E_{[I}{}^a
E_{J]}{}^b\left(A_a{}^I-\Gamma _a{}^I\right) \left(A_b{}^J
-\Gamma _b{}^J\right)\approx 0\,,\label{Asht3}\\
\end{eqnarray}
representing (\ref{Asht1}) the Gauss law, (\ref{Asht2}) the vector
constraint and (\ref{Asht3}) the scalar constraint respectively,
with both the covariant derivative
\begin{equation}
\nabla _a{} E_I{}^a :=\partial _a{} E_I{}^a +\epsilon
_{IJ}{}^KA_a^J E_K{}^a\,,\label{triadcovfiff}
\end{equation}
and the $SO(3)$ field strength
\begin{equation}
F_{ab}^I := 2\,\partial _{[a}A_{b]}^I +\epsilon
^I{}_{JK}\,A_a^J\,A_b^K\,,\label{fieldstrengthdef}
\end{equation}
built with the Barbero $SO(3)$ connection (\ref{Ashtekbeta3}),
being $\epsilon _{IJK}$ the $SO(3)$ Levi-Civita tensor. Since
(\ref{Asht1})--(\ref{Asht3}) are all of them first class
constraints \cite{Dirac50}, they behave as generators of
symmetries, corresponding (\ref{Asht1}) to the internal $SO(3)$
group, (\ref{Asht2}) to three-dimensional diffeomorphisms and
(\ref{Asht3}) to reparametrization invariance.
\bigskip

\section{\bf Ashtekar type variables from PGT}

We start making a brief comment on the relationship existing
between Ashtekar type formulations \cite{Ashtekar:1986yd}
\cite{Ashtekar:1987gu} \cite{Jacobson:1988yy}
\cite{Wallner:1990ng} \cite{Wallner:1992pj} and those alternative
realizations of GR as a gauge theory --very close to standard
Yang-Mills theories-- mainly due to Hehl and coworkers
\cite{Hehl:1976kj} \cite{Gronwald:1995em} \cite{Hehl:1995ue}
\cite{Tresguerres:2000qn} \cite{Tresguerres:2002uh}. The
correspondence between both approaches was extensively studied by
us in \cite{rrdph}. Here we summarize the basic points.

Hamiltonian treatments of GR based on any of the two mentioned
gauge-theoretical versions of gravity reveal to be transformable
into each other provided one identifies the Ashtekar variables
with certain combinations of quantities taken from the alternative
approach, as discussed below and shown in (\ref{Ashtekbeta2}). The
seemingly different symmetries involved --the $SO(3)$ internal
Ashtekar symmetry {\it versus}, say, the local Poincar\'e symmetry
of PGT-- are conciliated by the fact that nonlinear realizations
of groups allow to build actions being invariant under a given
group but manifestly invariant only under a subgroup of the total
gauge group. Accordingly, gravitational actions formulated
exclusively in terms of $SO(3)$--symmetric quantities can be
simultaneously invariant under a wider local spacetime symmetry
not explicitly displayed.

Let us consider the nonlinear PGT building blocks presented in the
Appendix. We are going to show how to get the Ashtekar
reformulation of the corresponding gravitational equations; in
particular of the Hamiltonian constraints. Since a homogeneous
notation is needed in order to compare the Ashtekar equations with
those built from the PGT quantities (\ref{PGT8})--(\ref{PGT11}),
we proceed to translate (\ref{Asht1})--(\ref{Asht3}) into the
language of exterior calculus\footnote{For what follows, an
important observation is in order. Due to the fact that we will
deal with constraints, one must have in mind that although we use
the same notation as in the Appendix, where the quantities
considered are previous to any foliation of spacetime, here we
only consider the spatial parts of such quantities. The latter
ones are defined on the spatial slices resulting from the
foliation with respect to the time vector $\hat{e}_0={1\over
N}\,\partial _\tau -N^a\,\partial _a$, having imposed the time
gauge in such a way that the time component of the tetrads reduces
to $\hat{\vartheta}^0 = N d\tau$.}. The Barbero-Ashtekar
connection (\ref{Ashtekbeta3}), rewritten as a 1-form $A^I=dx^a
A_a^I$, becomes
\begin{equation}
A^I :=\Gamma ^I +\beta\,K^I\,,\label{Ashtekbeta2}
\end{equation}
where, in general, the Immirzi parameter $\beta$ is a complex
number to be fixed. The link to PGT follows from identifying
(\ref{Ashtekbeta2}) as a combination of the real nonlinear PGT
connection fields (\ref{PGT10}) and (\ref{PGT11}), being $\Gamma
^I$ the connection associated to the $SO(3)$ subgroup of the
Lorentz group\footnote{$\Gamma ^I$ is expressible in terms of
frame and coframe fields as $\Gamma ^I (e)$ when torsion is
absent.}, while $K^I$ is the nonlinear connection of the boost
subgroup of the Lorentz group. (Notice that this gauge-theoretical
explanation of $K^I$ as a Poincar\'e quantity is compatible with
its usual interpretation as the extrinsic curvature). According to
(\ref{PGT15}) and (\ref{PGT14}), $\Gamma ^I$ transforms as an
$SO(3)$ connection and the nonlinear connection $K^I$ as an
$SO(3)$ tensor respectively, so that the composite Ashtekar type
variable (\ref{Ashtekbeta2}) --in general a complex quantity--
behaves as a modified $SO(3)$ connection in which both the $SO(3)$
part and and the boost part of the Lorentz connection of PGT are
comprised.

The interpretation of the Ashtekar formalism in terms of
Poincar\'e quantities is completed by identifying the coframes as
the nonlinear translational connections (\ref{PGT8}),
(\ref{PGT9})\footnote{Here we suppress the hat over the $SO(3)$
quantities present in the Appendix, where it is necessary in order
to distinguish them from the Lorentz ones.}, being in particular
(\ref{PGT9}) taken to be indistinguishable from the triads built
from the Ashtekar triad components $e_a{}^I$ as $\vartheta
^I:=dx^a e_a{}^I$, see (\ref{triad}), and transforming as an
$SO(3)$ covector, as shown in (\ref{PGT13}). (In order to
understand also the role played by the time component
(\ref{PGT8}), one should consider the whole four-dimensional
formalism previous to the foliation. The interested reader is
referred to our papers \cite{Lopez-Pinto:1997aw} \cite{rrdph}.)

The field strength built from (\ref{Ashtekbeta2}) is the 2-form
\begin{eqnarray}
\nonumber F^I :=&& d\,A^I +{1\over2}\,\epsilon ^I{}_{JK}\,A^J\wedge A^K\\
\nonumber =&&{\cal R}^I +{1\over 2}(\,1+\beta ^2\,)\,\epsilon
^I{}_{JK} K^J\wedge K^K +\beta\,
{\buildrel\Gamma\over{\nabla}}\,K^I\,,\label{betafieldstrbis}\\
\end{eqnarray}
whose components as given by $F^I = {1\over 2}\,F_{ab}^I
\,dx^a\wedge dx^b$ are identical with (\ref{fieldstrengthdef}). In
(\ref{betafieldstrbis}) we introduced the $SO(3)$ part of the
Poincar\'e curvature (\ref{PGT17}), defined as
\begin{equation}
{\cal R}^I :=\,d\,\Gamma ^I +{1\over2}\,\epsilon ^I{}_{JK}
\,\Gamma ^J\wedge \Gamma ^K -{1\over2}\,\epsilon
^I{}_{JK}\,K^J\wedge K^K\,,\label{App2.23}
\end{equation}
as much as the boost part of such curvature, that is
\begin{equation}
{\buildrel\Gamma\over{\nabla}}\,K^I :=\,d\,K^I +\epsilon
^I{}_{JK}\,\Gamma ^J\wedge K^K\,.\label{App2.22}
\end{equation}
In the following we simplify things by fixing $\beta =i$ in
(\ref{Ashtekbeta2}), so that the latter becomes a standard complex
Ashtekar variable \cite{Barbero:1995ap} \cite{Randono:2005hb} and
(\ref{betafieldstrbis}) reduces to
\begin{equation}
F^I ={\cal R}^I
+i\,{\buildrel\Gamma\over{\nabla}}\,K^I
\,.\label{Ashtek2b}
\end{equation}
We do not enter the discussion on the physical relevance of the
value assigned to the Barbero--Immirzi parameter
\cite{Barbero:1995ap} \cite{Immirzi:1996dr} \cite{Randono:2005hb}
\cite{Kodama:1990sc} \cite{Ashtekar:2000eq}. Let us merely mention
that at the classical level Einstein's equations have the same
physical content for any value of $\beta$, being different choices
of such value related to each other by means of canonical
transformations. The Immirzi quantization ambiguity concerns
quantum gravity, where distinct phase spaces corresponding to
different values of $\beta$ correspond to unitarily inequivalent
quantum theories \cite{Ashtekar:2000eq}. However, the possible
role of $\beta$ as a fundamental constant in quantum gravity, to
be measured empirically, seems to be not well understood
\cite{Samuel:2000ue}; and in any case our choice could be changed
if necessary.

We end this section using (\ref{Ashtek2b}) and of the triads
(\ref{PGT9}) to translate the constraints
(\ref{Asht1})--(\ref{Asht3}) into the language of exterior
calculus. The Gauss constraint (\ref{Asht1}) can be rewritten as
\begin{equation}
{\cal G}_I = e\,\,{}^\#\left(\nabla\eta _I\right)\approx\,
0\,,\label{Gauss}
\end{equation}
in terms of the eta-basis 2-form defined as
\begin{equation}
\eta _{_I} :={1\over {2!}}\,\epsilon _{_{IJK}}\,\vartheta
^J\wedge\vartheta ^K ={1\over {2!}}\,\epsilon _{_{IJK}}\,
e_a{}^J\,e_b{}^K\,dx^a\wedge dx^b\,.\label{eta1def}
\end{equation}
The symbol $^\#$ stands for the Hodge dual in three dimensions;
see for instance \cite{Lopez-Pinto:1997aw}. (In particular,
(\ref{eta1def}) is the Hodge dual of the triad, that is $\eta
_I\equiv {}^\#\vartheta _I$.) On the other hand, the vector
constraint (\ref{Asht2}) becomes
\begin{equation}
dx^a {\cal V}_a :=dx^a E_I{}^b F^I_{ab} =e\,{}^\#\left(\,\vartheta
_I\wedge{}^\# F^I\,\right)\approx\,0\,,\label{vector}
\end{equation}
where the three-dimensional Hodge dual of $F^I = {1\over
2}\,F_{ab}^I \,dx^a\wedge dx^b ={1\over 2}\,(e_K\rfloor e_J\rfloor
F^I\,)\,\vartheta ^J\wedge\vartheta ^K$ reads ${}^\# F^I ={1\over
2}\,(e_K\rfloor e_J\rfloor F^I\,)\,\epsilon ^{JK}{}_L\vartheta
^L$. Analogously, the scalar constraint (\ref{Asht3}) (with the
chosen value $\beta =i$) takes the form
\begin{equation}
{\cal S} :=\,\epsilon _I{}^{JK} E_J{}^a E_K{}^b F^I_{ab}=
2\,e^2\,{}^\#\left(\,\vartheta _I\wedge
F^I\,\right)\approx\,0\,,\label{scalar}
\end{equation}
where we made use of the fact that the Hodge dual of the
three-form $dx^a\wedge dx^b\wedge dx^c$ is identical with the
three-dimensional Levi-Civita tensor density $\eta ^{abc}$, that
is ${}^\#(\,dx^a\wedge dx^b\wedge dx^c\,) =\eta ^{abc} =\epsilon
^{IJK}\,e_I{}^a\,e_J{}^b\,e_K{}^c $. In view of (\ref{Gauss}),
(\ref{vector}) and (\ref{scalar}), we can replace the constraints
(\ref{Asht1})--(\ref{Asht3}) by their exterior calculus
formulation
\begin{eqnarray}
\nabla\eta _I &\approx &\, 0\qquad {\rm (GAUSS)}\,,\label{extcG}\\
\vartheta _I\wedge{}^\# F^I &\approx &\,0\qquad {\rm (VECTOR)}\,,\label{extcV}\\
\vartheta _I\wedge F^I &\approx &\,0\qquad {\rm (SCALAR)}
\,.\label{extcS}
\end{eqnarray}
The tortuous way followed here to get the form
(\ref{extcG})--(\ref{extcS}) of the constraints is justified by
the attempt to make evident the relationship between PGT and the
Ashtekar formalism. However, the interested reader is referred to
\cite{rrdph}, where we deduced such constraints directly from a
Poincar\'e invariant gravitational action.
\bigskip

\section{\bf Previous partial solutions of the constraints}

A new solution for the system (\ref{extcG})--(\ref{extcS}) --to be
added to any other possible solution-- will be studied in next
section, fulfilling both the Gauss law (\ref{extcG}) and the
vector constraint (\ref{extcV}), leaving only the scalar
constraint (\ref{extcS}) unsolved. The novelty of that solution,
given by Eq. (\ref{solution1}) below expressing the extrinsic
curvature in terms of the three-metric, is emphasized by comparing
it with previous results. Let us recall here in particular the one
due to Capovilla, Dell and Jacobson \cite{Capovilla:1989ac}, which
formally solves the vector and scalar constraints (\ref{extcV})
and (\ref{extcS}) respectively, and the related Barbero proposal
\cite{Barbero:1994tw} solving the Gauss law (\ref{extcG}) and the
scalar constraint (\ref{extcS}). We review both approaches in a
simple reformulation adapted to the notation introduced in
previous section.

In essence, the solution of the type proposed by Capovilla et al.
\cite{Capovilla:1991kx} rests on a suitable form to depict the
components of the field strength (\ref{Ashtek2b}), namely
\begin{equation}
F^I =\Psi ^{IJ}\eta _J\,,\label{Capovilla}
\end{equation}
with $\Psi ^{IJ}$ as a 3$\times$3 complex matrix and $\eta _I$ as
the 2-form given by (\ref{eta1def}). The vector constraint
(\ref{extcV}) then turns into
\begin{equation}
\vartheta _I\wedge{}^\# F^I =\Psi ^{IJ}\vartheta _I\wedge\vartheta
_J\approx\,0\,,\label{extcV2}
\end{equation}
while the scalar constraint (\ref{extcV}) becomes
\begin{equation}
\vartheta _I\wedge F^I =\Psi ^{IJ}\vartheta _I\wedge\eta _J =\Psi
_I{}^I\eta \approx\,0\,.\label{extcS2}
\end{equation}
(In (\ref{extcS2}) we made use of the identity $\vartheta ^I\wedge
\eta _J =\delta ^I_J\,\eta$, with $\eta :={1\over{3!}}\,\epsilon
_{_{IJK}}\,\vartheta ^I\wedge\vartheta ^J\wedge\vartheta ^K$ as
the volume element.) From (\ref{extcV2}) and (\ref{extcS2}) one
trivially reads out that both the vector and the scalar constraint
are satisfied by any symmetric traceless matrix $\Psi ^{IJ}$.
However, (\ref{extcG}) is not automatically fulfilled. Instead, it
gives rise to the condition
\begin{equation}
\left[ \nabla (\Psi ^{-1})_{IJ}\right]\wedge
F^J\approx\,0\,,\label{extcG2}
\end{equation}
involving three differential equations to be solved. This makes
the solution of Capovilla et al. unsatisfactorily incomplete, not
to speak about its rather formal character.

An alternative to the previous solution was proposed by Barbero
\cite{Barbero:1994tw}. Let us briefly revise it, suitably
accommodated to our notation. Making use of (\ref{Ashtek2b}), and
taking into account the identity
${\buildrel\Gamma\over{\nabla}}\vartheta ^I\equiv 0$ following
from (\ref{SO3conn}), (\ref{SO3connbis}) --with
${\buildrel\Gamma\over{\nabla}}$ as given by (\ref{App2.21c})--,
we separate equations (\ref{extcG})--(\ref{extcS}) into their real
and imaginary parts as
\begin{eqnarray}
\nonumber 0&\approx & \nabla\eta _I
={\buildrel\Gamma\over{\nabla}}\eta _I +i\,\epsilon _{IJ}{}^K
K^J\wedge\eta _K\\
&&\qquad = i\,\vartheta _I\wedge\vartheta _J\wedge K^J\,,\label{extcGreim}\\
0&\approx & \vartheta _I\wedge{}^\# F^I = \vartheta _I\wedge{}^\#
{\cal R}^I +i\,\vartheta
_I\wedge{}^\#{\buildrel\Gamma\over{\nabla}}\,K^I\,,\label{extcVreim}\\
0&\approx & \vartheta _I\wedge F^I =\vartheta _I\wedge {\cal R}^I
-i\,d\left(\,\vartheta _I\wedge K^I\,\right)\,.\label{extcSreim}
\end{eqnarray}
In analogy with the complex matrix $\Psi ^{IJ}$ in
(\ref{Capovilla}), we propose a real matrix $\Phi ^{IJ}$ such that
\begin{equation}
{\cal R}^I =\Phi ^{IJ}\eta _J\,.\label{FB1}
\end{equation}
By introducing the notation
\begin{eqnarray}
{\buildrel\Gamma\over {F^I}}&:=&\,d\,\Gamma ^I
+{1\over2}\,\epsilon ^I{}_{JK} \,\Gamma ^J\wedge \Gamma
^K\,,\label{FB1bis}\\
{}^\# S^I&:=& \,\epsilon ^I{}_{JK} \,K^J\wedge K^K
\,,\label{FB1bisbis}
\end{eqnarray}
so that (\ref{App2.23}) decomposes into
\begin{equation}
{\cal R}^I =\,{\buildrel\Gamma\over {F^I}} -{1\over2}\,\,{}^\# S^I
\,.\label{FB2}
\end{equation}
From (\ref{FB2}) with (\ref{FB1}) we get
\begin{equation}
S^I = 2\,(\,{}^\# {\buildrel\Gamma\over {F^I}} -\Phi
^{IJ}\vartheta _J\,)\,,\label{FB3}
\end{equation}
while on the other hand the inversion of (\ref{FB1bisbis}) yields
\begin{equation}
K^I ={}^\#\left({1\over{2\sqrt{2S}}}\,\epsilon ^I{}_{JK}
\,S^J\wedge S^K\,\right)\,,\label{FB4}
\end{equation}
being $S$ the determinant $S:={1\over{3!}}\,\eta ^{abc}\epsilon
_{IJK}\,S^I_a\,S^J_b\,S^K_c\,$. The condition for both
(\ref{extcGreim}) and the imaginary part of (\ref{extcSreim}) to
vanish simultaneously is $\vartheta _I\wedge K^I =0$, which by
invoking (\ref{FB4}) gives rise to
\begin{equation}
0=\vartheta _I\wedge K^I =-{1\over{\sqrt{2S}}}\,\epsilon
^I{}_{JK}\,\left(\,e_I\rfloor S^J\,\right)\,{}^\#
S^K\,.\label{FB6}
\end{equation}
Replacing (\ref{FB3}) into (\ref{FB6}) taking into account that
$\epsilon ^I{}_{JK}\,(\,e_I\rfloor {}^\# {\buildrel\Gamma\over
{F^J}} \,) =-\,{}^\#
(\,{\buildrel\Gamma\over{\nabla}}{\buildrel\Gamma
\over{\nabla}}\vartheta _K\,)\equiv 0$, condition (\ref{FB6})
reduces to $\Phi _{[IJ]}=0$. As a byproduct, the symmetric $\Phi
_{IJ}$ ensuring the vanishing of (\ref{extcGreim}) and of the
imaginary part of (\ref{extcSreim}) also cancels the real part of
(\ref{extcVreim}), since the latter, with (\ref{FB1}), reads
$\vartheta _I\wedge{}^\# {\cal R}^I = \Phi ^{IJ}\,\vartheta
_I\wedge\vartheta _J $. On the other hand, the real part of
(\ref{extcSreim}) reduces to $\vartheta _I\wedge {\cal R}^I =\Phi
^{IJ}\,\vartheta _I\wedge\eta _J =\Phi _I{}^I\,\eta $, thus
vanishing for $\Phi _I{}^I =0$. So, any symmetric traceless matrix
$\Phi _{IJ}$ solves the system
(\ref{extcGreim})--(\ref{extcSreim}) up to the imaginary part of
the vector constraint (\ref{extcVreim}).
\bigskip

\section{Three-metric dependent solution for the extrinsic curvature}

The approach proposed by us mainly differs from the previously
considered ones in that we pay attention to the connections --that
is, to the fundamental variables-- rather than to the field
strengths built with them. Actually, not the components of $F^I$
or of ${\cal R}^I$ as in previous section but those of the
nonlinear boost connection $K^I$ (identical with the ADM extrinsic
curvature) play the relevant role as the quantities to be
determined.

Starting from the constraints in their form
(\ref{extcGreim})--(\ref{extcSreim}), we find that the Gauss law
(\ref{extcGreim}) is trivially fulfilled for the extrinsic
curvature $K_{IJ}:=e_I\rfloor K_J$ a symmetric matrix. This result
can alternatively be obtained from (\ref{Asht1}). Indeed, by
replacing in it (\ref{Ashtekbeta3}), due to the fact that the
$\beta$-independent part of the covariant derivative vanishes
identically, we get
\begin{equation}
\nabla _a{} E_I{}^a = \beta\,\epsilon _{IJ}{}^M K_a{}^J
E_M{}^a\approx\,0\,,\label{Gauss2}
\end{equation}
which is automatically satisfied by any symmetric matrix
$K_M{}^J{}:= K_a{}^J e_M{}^a$. As a corollary of $K_{[IJ]}=0$, the
real contribution of (\ref{extcVreim}) automatically vanishes, as
it is easily seen by explicitly displaying
\begin{equation}
\vartheta _I\wedge{}^\# {\cal R}^I =e_I\rfloor\left(
{\buildrel\Gamma\over{\nabla}}{\buildrel\Gamma\over{\nabla}}\vartheta
^I +K^I\wedge K_J\wedge\vartheta
^J\right)\approx\,0\,,\label{firstvector3}
\end{equation}
where ${\buildrel\Gamma\over{\nabla}}\vartheta ^I$ stands for the
vanishing spatial contribution to torsion, while $\vartheta
_J\wedge K^J$ is zero in view of the symmetry of $K_{IJ}$. For the
same reason, the imaginary contribution of (\ref{extcSreim}) is
also null. Consequently, Eqs. (\ref{extcVreim}) and
(\ref{extcSreim}) reduce respectively to
\begin{eqnarray}
\nonumber 0 &\approx &\vartheta _I\wedge{}^\# F^I = i\,\vartheta
_I\wedge{}^\#{\buildrel\Gamma\over{\nabla}}\,K^I\\
\nonumber &=&-{i\over 2}\,\epsilon
^{IJK}e_M{}^a{\buildrel\Gamma\over{\nabla}} _a\left(\,K_K{}^M
-\delta _K^M\,K_N{}^N\right) \vartheta
_I\wedge\vartheta _J\,,\label{vector5}\\
\end{eqnarray}
and
\begin{eqnarray}
\nonumber 0 &\approx &\vartheta _I\wedge F^I =\vartheta _I\wedge
{\cal R}^I\\
&=&-{1\over 2}\,\eta\left[\,{}^{(3)}R -K_I{}^J K_J{}^I+K_I{}^I
K_J{}^J\,\right]\,,\label{scalarexplicit}
\end{eqnarray}
with ${}^{(3)}R$ as the Riemannian scalar curvature built from the
three-metric $q_{\,ab}:=\delta _{IJ}\,e_a{}^I e_b{}^J\,$.
Comparing (\ref{vector5}), (\ref{scalarexplicit}) with
(\ref{extcV2}), (\ref{extcS2}), the explicit form of $\Psi
^{[IJ]}$ and $\Psi _I{}^I$ follows in terms of $K_{IJ}$. The
transition to the standard ADM formalism can be easily performed.
Making use of (\ref{vector}) with (\ref{vector5}), we calculate
\begin{eqnarray}
\nonumber {i\over e}\,{\cal V}_a &=&i\,\epsilon _{IJK}\Psi
^{IJ}e_a{}^K = e_a{}^K e_M{}^b\,{\buildrel\Gamma\over{\nabla}}
_b\left(\,K_K{}^M -\delta _K^M\,K_N{}^N\right)\\
&=& D_b\left(\,K_a{}^b -\delta _a^b\,K_c{}^c\right)\approx\,0
\,,\label{momconstr}
\end{eqnarray}
being $K_{ab}:=K_a{}^I e_{bI}$, while $D_a$ is the covariant
derivative associated with the 3-metric $q_{ab}$, built with the
ordinary Christoffel symbol (\ref{Christoffel3}). In the vector
constraint written as (\ref{momconstr}) we recognize the standard
form of the momentum constraint in ADM variables
\cite{Alvarez:1988tb} with the extrinsic curvature playing the
role of the momentum. On the other hand, from (\ref{scalar}) with
(\ref{scalarexplicit}), and in view of the invariance of the
expressions involved, one finds
\begin{equation}
-{1\over{e^2}}\,{\cal S} =-2\,\Psi _I{}^I =\,{}^{(3)}R -K_a{}^b
K_b{}^a+K_a{}^a K_b{}^b\approx\,0\,.\label{scalconstr}
\end{equation}
So, from (\ref{momconstr}) and (\ref{scalconstr}) we recover the
standard ADM constraints \cite{Alvarez:1988tb}, which for clarity
we display separately as
\begin{eqnarray}
D_b\left(\,K_a{}^b -\delta _a^b\,{\rm tr} K\right)&\approx &\,0
\,,\label{momconstrbis}\\
R -{\rm tr}(K^2) +({\rm tr} K)^2 &\approx
&\,0\,.\label{scalconstrbis}
\end{eqnarray}
(In (\ref{scalconstrbis}) and in the following we use the
simplified notation $R:={}^{(3)}R$.) The variables concerned in
(\ref{momconstrbis}) are the matrix $K_{ab}$ (in the following we
refer to it simply as $K$) as much as the non explicitly displayed
three-metric $q_{ab}$ present in the Christoffel symbols of the
covariant derivative. Being $K$ symmetric, in each point of the
spacetime manifold it is diagonalizable by means of an orthogonal
transformation, so that its eigenvalues arranged in the diagonal
matrix turn out to be the relevant variables, while the three
degrees of freedom of the diagonalizing orthogonal matrix become
absorbed into the three-metric and thus --let us say--
geometrized. As it is well known, from a $3\times 3$ matrix $K$ it
is possible to build three invariants, namely its trace ${\rm tr}
K$, the trace ${\rm tr}(K^2)$ of the matrix square, and the
determinant $\det K$, being such invariants related to the matrix
$K$ by the characteristic equation
\begin{equation}
K^3 - ({\rm tr} K)\,K^2 +{1\over 2}\,[ ({\rm tr} K)^2 -{\rm
tr}(K^2)\,]\,K - I \det K =0\,,\label{chareq}
\end{equation}
see \cite{Capovilla:1991kx}. Being the invariants expressible in
terms of the matrix eigenvalues, from now on we pay attention to
the invariants rather than to the eigenvalues.

One can diminish the number of invariants involved in
(\ref{momconstrbis}), (\ref{scalconstrbis}) by imposing the zero
trace assumption ${\rm tr} K =0$, the latter constituting the
usual slicing condition used as a gauge fixing of the
reparametrization invariance \cite{Alvarez:1988tb}. Accordingly,
(\ref{scalconstrbis}) reduces to
\begin{equation}
{\rm tr}(K^2) \approx \,R\,,\label{scalconstrbisbis}
\end{equation}
thus determining a further invariant, namely ${\rm tr} (K^2)$, in
terms of the scalar curvature; that is, in terms of a quantity
built from the three-metric $q_{ab}$ of the underlying Riemannian
space and from their derivatives up to second order. So, only
$\det K$ remains as an unknown function to be determined from the
differential condition (\ref{momconstrbis}) in its simplified form
\begin{equation}
D_b\,K_a{}^b \approx\,0\,,\label{momconstrbisbis}
\end{equation}
which (for diagonalized $K$) relates the first derivatives of the
remaining invariant $\det K$ with the invariant itself and with
several geometrical objects involving the metric and its
derivatives. As a result, as far as one primarily looks for
solutions of $K$ being functionals of arbitrary metrics (other
solutions will be considered below), $\det K$ has to be
expressible as an invariant built with the metric tensor $q_{ab}$
and its derivatives up to third order. The reason for it is that
the first derivative of the invariant ${\rm tr}(K^2)$, implicitly
present in (\ref{momconstrbisbis}) through
(\ref{scalconstrbisbis}), yields a derived scalar curvature $R$,
the latter comprising up to second derivatives of the metric.

Having in mind the previous discussion, let us follow an indirect
way to find a matrix $K$ satisfying all the required features.
First of all we parametrize such matrix as
\begin{equation}
K^{ab}=\,\eta ^{acd} D_c\,S_d{}^b\,,\label{ansatz1}
\end{equation}
in terms of the (derived) matrix $S_d{}^b$ to be determined. The
latter matrix must be chosen so that it guarantees that, as
studied above, $K^{ab}$ is symmetric and traceless, satisfying in
addition equation (\ref{momconstrbisbis}). From (\ref{ansatz1})
follows
\begin{equation}
K^{ab}\,\eta _{abc} =D_b (\,S_c{}^b -\delta
_c^b\,S_d{}^d\,)\,,\label{ansatz2}
\end{equation}
showing that the symmetry condition on (\ref{ansatz1}) requires
the vanishing of the r.h.s. of (\ref{ansatz2}). On the other hand,
(\ref{ansatz1}) is trivially traceless provided $S_{db}$ is a
symmetric matrix. Finally, we find the divergence of
(\ref{ansatz1}) to be
\begin{equation}
D_b K_a{}^b= \,\eta _a{}^{bc} R_{bd} S_c{}^d + \eta _a{}^{bc}D_b
D_d S_c{}^d\,,\label{ansatz3}
\end{equation}
where we made use of the formula (\ref{solution3}) below, holding
in three dimensions. Due to the covariance of $K$, we make the
ansatz --in the particular case considered here that $K$ is taken
to be a functional of an arbitrary three-metric $q_{ab}$ and of at
most its third derivatives-- that the matrix $S_d{}^b$ in
(\ref{ansatz1}) has to be built from the Riemann tensor and its
contractions. We know that in three dimensions no Weyl tensor
exists \cite{Weinberg}, so that the Riemann tensor $R_{cbd}{}^f
:=2\,\Bigl(\,\partial _{[c}\Gamma _{b]d}{}^f +\Gamma
_{[ch}{}^f\Gamma _{b]d}{}^h\,\Bigr)$ can be decomposed in terms of
the Ricci tensor $R_{cd}:=R_{fcd}{}^f$ and of the scalar curvature
$R:=g^{cd} R_{cd}\,$ as
\begin{equation}
R_{cbd}{}^f =-2\,q_{d[c}R_{b]}{}^f +2\,\delta _{[c}^f R_{b]d}
-\delta _{[c}^f q_{b]d} R\,.\label{solution3}
\end{equation}
In particular we choose for $S_d{}^b$ the combination
\begin{equation}
S_d{}^b = k\,R_d{}^b + k'\,\delta _d^b\,R\,,\label{ansatz4}
\end{equation}
which is symmetric as required and satisfies (\ref{ansatz3})
automatically (recall the identity $D_d G_c{}^d\equiv 0$, with
$G_c{}^d$ as the Einstein tensor). The remaining condition of
vanishing (\ref{ansatz2}) enforces us to fix $k'=-k/4$, so that
(\ref{ansatz1}) with (\ref{ansatz4}) transforms into
\begin{equation}
K^{ab}=k\,\eta ^{acd} D_c\left( R_d{}^b -{1\over 4}\,\delta
_d^b\,R\,\right)\,,\label{solution1}
\end{equation}
satisfying all the requirements stipulated for $K$. In the r.h.s.
of (\ref{solution1}) we recognize the Cotton-York tensor (also
called Bach tensor) built from the Cotton tensor in 3-dimensional
space, see eq.(97) of \cite{Garcia:2003bw}. The five degrees of
freedom of the symmetric traceless exterior curvature thus become
expressed --through the five independent components of the
Cotton-York tensor-- in terms of any three-metric $q_{ab}$ without
additional integrability conditions.

Notice that (\ref{scalconstrbisbis}) is not completely foreign to
our main result (\ref{solution1}). Actually --as we will
immediately see-- other solutions are possible for the divergence
condition (\ref{momconstrbisbis}). In the discussion preceding our
concrete proposal, it was precisely (\ref{scalconstrbisbis}) that
played a crucial role in justifying to consider a contribution to
$K$ built exclusively as a functional of an arbitrary
three-metric, by reducing the invariant function ${\rm tr}(K^2)$
to the $q_{ab}$--dependent scalar curvature. (Recall that by also
fixing ${\rm tr} K =0$, the only remaining invariant $\det K$ and
its derivatives became related solely to functions of the
three-metric.) Thus, the general validity of (\ref{solution1})
rests on having taken simultaneously into account
(\ref{scalconstrbisbis}) as a consistence condition. In other
words, (\ref{solution1}) can be taken as a general metric
dependent solution as far as it refers to the system
(\ref{momconstrbisbis}), (\ref{scalconstrbisbis}) rather than
merely to (\ref{momconstrbisbis}).

Observe that for three-metrics such that $R=0$, no independent
dynamics for $K_{ab}$ exists any more if one requires the
eigenvalues of $K$ to be real. Indeed, with $R=0$,
(\ref{scalconstrbisbis}) implies ${\rm tr}(K^2)=0$. Having already
assumed ${\rm tr} K =0$, the characteristic equation
(\ref{chareq}) reduces to
\begin{equation}
K^3 - I \det K =0\,.\label{modchareq}
\end{equation}
The discriminant involved in the solution of (\ref{modchareq}) is
positive, so that the eigenvalues of $K$ consist of one real and
two complex conjugate roots, thus leaving $K_{ab} =0$ as the only
physically admissible extrinsic curvature.

\section{Additional solutions of $K_{ab}$}

The natural question to be asked now is about the existence of
nontrivial $K$'s which, although necessarily linked to a metric
with $R\neq 0$, are nevertheless irreducible to the functional
(\ref{solution1}) of such metric. We are going to prove that, in
fact, (\ref{solution1}) does not exhaust all possible solutions of
the constraints (\ref{scalconstrbisbis}) and
(\ref{momconstrbisbis}).

Due to the interplay existing between $K$ and the three-metric, it
is illustrative to consider isotropic metrics, for which
(\ref{solution1}) vanish, guaranteeing that any possible $K\neq 0$
must be independent of (\ref{solution1}). The general isotropic
metric is given by its nonvanishing components
\begin{equation}
q_{11}={1\over{\phi}}\,,\quad q_{22}=r^2\,,\quad q_{33}=r^2\,\sin
^2\theta\,,\label{explicitmetric}
\end{equation}
where $\phi =\phi (r)$. The corresponding components of the
Christoffel connection (\ref{Christoffel3}) read
\begin{eqnarray}
\Gamma _{11}^1 &=&-{{\partial _r\phi}\over{2\phi}}\,,\quad \Gamma
_{22}^1 =-r\phi\,,\quad \Gamma _{33}^1 =-r\phi\,\sin
^2\theta\,,\nonumber\\
\Gamma _{12}^2 &=& \Gamma _{21}^2 = {1\over r}\,,\quad \Gamma
_{33}^2
=-\sin\theta\,\cos\theta\,,\nonumber\\
\Gamma _{13}^3 &=& \Gamma _{31}^3 ={1\over r}\,,\quad \Gamma
_{23}^3 = \Gamma _{32}^3 =\cot\theta\,,\label{explicitChris}
\end{eqnarray}
the Ricci tensor $R_{ab} :=2\,\Bigl(\,\partial _{[a}\Gamma
_{c]b}{}^c +\Gamma _{[ad}{}^c\Gamma _{c]b}{}^d\,\Bigr)$ calculated
from (\ref{explicitChris}) is diagonal, with
\begin{eqnarray}
R_{11}&&={{\partial _r\phi}\over{r\phi}}\,,\nonumber\\
R_{22}&&={1\over{2r}}\,\partial _r\left[\,r^2\left(\,\phi
-1\,\right)\,\right]\,,\nonumber\\
R_{33}&&={{\sin ^2\theta}\over{2r}}\,\partial
_r\left[\,r^2\left(\,\phi -1\,\right)\,\right]\,,\label{Ricci}
\end{eqnarray}
and the scalar curvature reads
\begin{equation}
R := q^{ab}\,R_{ab} ={2\over{r^2}}\,\partial
_r\left[\,r\left(\,\phi -1\,\right)\,\right]\,.\label{explicitR}
\end{equation}
It is trivial to check, by replacing (\ref{Ricci}) and
(\ref{explicitR}), that the solution (\ref{solution1}) of $K_{ab}$
reducible to the three-metric vanishes for any isotropic metric.

In order to simplify calculations, let us rewrite the matrix
$K_a{}^b$ (with general indices) as
\begin{equation}
K_a{}^b = e_a{}^M\,K_M{}^N\,e^b{}_N\,,\label{prediagonalization}
\end{equation}
in terms of the symmetric matrix with internal indices $K_M{}^N$,
which can be diagonalized with the help of an orthogonal matrix
$O_M{}^I$ as $K_M{}^N = O_M{}^I\,\Delta _I{}^J\,O^N{}_J$ with
$\Delta _I{}^J$ diagonal. We write (\ref{prediagonalization}) as
\begin{equation}
K_a{}^b =\hat{e}_a{}^I\,\Delta
_I{}^J\,\hat{e}^b{}_J\,,\label{diagonalization}
\end{equation}
by introducing the redefined ${\it dreibein}$ $\hat{e}_a{}^I
:=e_a{}^M\,O_M{}^I$. Accordingly, (\ref{momconstrbisbis}) build
with (\ref{Christoffel3}) transforms into
\begin{equation}
0= D_b\,K_a{}^b =\partial _b\rfloor D\,K_a{}^b =\left(
\hat{e}_I\rfloor\hat{D}\Delta _J{}^I\,\right)\,\hat{e}_a{}^J
\,,\label{reformulmomconstrbisbis}
\end{equation}
where the relation (\ref{SO3connbis}) was taken into account,
being
\begin{equation}
\hat{\Gamma}_{IJ}:=\hat{e}_{[\,I}\rfloor d\hat{\vartheta}_{J]}
-{1\over 2}\,\left(\hat{e}_I\rfloor\hat{e}_J\rfloor
d\hat{\vartheta}^K\,\right)\,\hat{\vartheta}_K
\,,\label{triadconn}
\end{equation}
with $\hat{\vartheta}^I := dx^a\,\hat{e}_a{}^I$, and
\begin{equation}
\hat{D}\Delta _J{}^I := d\,\Delta _J{}^I +
\hat{\Gamma}_K{}^I\,\Delta _J{}^K -\hat{\Gamma}_J{}^K\,\Delta
_K{}^I\,.\label{diffdef}
\end{equation}
In the following we fix the gauge of the $SO(3)$ symmetry by
choosing $O_M{}^I =\delta _M^I$, so that $\hat{e}_a{}^I =
e_a{}^I$. (Accordingly, from now on we suppress the hat over the
triads as much as over the connections.)

Eq.(\ref{reformulmomconstrbisbis}) shows the correspondence
between the alternative notations in terms of the three-metric or
of triads respectively. Then, instead of the metric components
(\ref{explicitmetric}) and the ordinary Christoffel symbols
(\ref{explicitChris}), one can introduce the triads
\begin{equation}
\vartheta ^r ={dr\over{\sqrt{\phi}}}\,,\quad \vartheta ^\theta
=r\,d\theta\,,\quad \vartheta ^\varphi
=r\,\sin\theta\,d\varphi\,,\label{triadcomps}
\end{equation}
and the corresponding connection 1-form (\ref{triadconn}), with
antisymmetric components
\begin{equation}
\Gamma _{r\theta} ={{\sqrt{\phi}}\over r}\,\vartheta
^\theta\,,\quad \Gamma _{r\varphi} ={{\sqrt{\phi}}\over
r}\,\vartheta ^\varphi\,,\quad \Gamma _{\theta\varphi}
={{\cot\theta}\over r}\,\vartheta
^\varphi\,.\label{triadconncomps}
\end{equation}
Making use of (\ref{triadcomps}) and (\ref{triadconncomps}), we
find the explicit form of (\ref{reformulmomconstrbisbis}) by
calculating
\begin{eqnarray}
e_I\rfloor D\Delta _r{}^I &=& {{\sqrt{\phi}}\over
r}\,\Bigl[\,\partial _r\left(\,r^2\Delta _r{}^r\,\right) -r
\left(\,\Delta _\theta{}^\theta +\Delta
_\varphi{}^\varphi\,\right)\,\Bigr] \,,\label{eqcomp1}\\
e_I\rfloor D\Delta _\theta{}^I &=& {1\over r}\,\Bigl[\,\partial
_\theta\Delta _\theta{}^\theta +\cot\theta \left(\,\Delta
_\theta{}^\theta -\Delta
_\varphi{}^\varphi\,\right)\,\Bigr] \,,\label{eqcomp2}\\
e_I\rfloor D\Delta _\varphi{}^I &=& {1\over
{r\sin\theta}}\,\partial _\varphi\Delta
_\varphi{}^\varphi\,.\label{eqcomp3}
\end{eqnarray}
We propose the suitable parametrization
\begin{eqnarray}
\Delta _r{}^r &=& \lambda\cos\omega\,,\label{paramet1}\\
\Delta _\theta{}^\theta &=& -{{\lambda}\over 2}\left(\,\cos\omega
-\sqrt{3}\,\sin\omega\,\right)\,,\label{paramet2}\\
\Delta _\varphi{}^\varphi &=& -{{\lambda}\over
2}\left(\,\cos\omega +\sqrt{3}\,\sin\omega\,\right)
\,,\label{paramet3}
\end{eqnarray}
for a (diagonalized) real traceless matrix, in terms of which
equations (\ref{reformulmomconstrbisbis}) with
(\ref{eqcomp1})--(\ref{eqcomp3}) transform into
\begin{eqnarray}
0&=& \partial _r\left(\,r^3\,\lambda\cos\omega\,\right)
\,,\label{firsteq}\\
0&=&\lambda\left[\,\partial _\theta\cos\omega
-{{\sqrt{3}}\over{\sin ^2\theta}}\,
\partial _\theta\left(\,\sin\omega\,\sin ^2\theta\,\right)\right]
\,,\label{secondeq}\\
0&=&\lambda\,\partial _\varphi\left(\,\cos\omega
+\sqrt{3}\,\sin\omega\,\right)\,.\label{thirdeq}
\end{eqnarray}
The choice (\ref{paramet1})--(\ref{paramet3}) has the virtue of
guaranteeing automatically ${\rm tr} K =0$, while $\det K ={1\over
4}\,\lambda ^3\cos\omega\left( 1-4\sin ^2\omega\right)$ and ${\rm
tr} (K^2) ={3\over 2}\,\lambda ^2 $. Replacing the last expression
as much as (\ref{explicitR}) into (\ref{scalconstrbisbis}), we get
\begin{equation}
{3\over 2}\,\lambda ^2 = {2\over{r^2}}\,\partial
_r\left[\,r\left(\,\phi -1\,\right)\,\right]
\,,\label{isotropscalconstr}
\end{equation}
implying that $\lambda$ exclusively depends on $r$.

Let us first consider the case $R=0$, noting that we are working
with a metric different from the trivial Euclidean one. As read
out from (\ref{explicitR}) and (\ref{isotropscalconstr}), $R=0$
corresponds to $\lambda =0$, so that
(\ref{firsteq})--(\ref{thirdeq}) are trivially fulfilled. On the
other hand, (\ref{isotropscalconstr}) reduces to $\partial
_r\left[\,r\left(\,\phi -1\,\right)\,\right] =0$, whose solution
reads
\begin{equation}
r\left(\,\phi -1\,\right) =-2m\,,\label{Schwarz1}
\end{equation}
with the integration constant suitably denoted as a mass in order
to reproduce the well known Schwarzshild function solution in
(\ref{explicitmetric}) or (\ref{triadcomps}). We read it out from
(\ref{Schwarz1}) to be
\begin{equation}
\phi= 1-{{2m}\over r}\,.\label{Schwarz2}
\end{equation}
As discussed above, $K_{ab}=0$. No $K$ exists in this case,
neither as (\ref{solution1}) nor as a different
--metric-independent-- contribution.

Less trivial is the case with $R\neq 0$, and thus with
$\lambda\neq 0$. Certainly, (\ref{solution1}) vanishes as for any
isotropic three-metric. However, a nontrivial $K$ not reducible to
(\ref{solution1}) can be found as follows. By expressing
(\ref{secondeq}) in the form
\begin{equation}
0=\,\sin\omega\,\partial _\theta\omega +\sqrt{3}\,\left(\,\partial
_\theta\sin\omega +2\,\sin\omega
\,\cot\theta\,\right)\,,\label{modsecondeq}
\end{equation}
it is clear that it is satisfied for $\sin\omega =0$. This is the
case we are going to consider.

But let us briefly show before what happens for $\sin\omega\neq
0$. By rewriting (\ref{firsteq}) and (\ref{secondeq}) respectively
as
\begin{equation}
\partial _r\omega =\cot\omega\,\partial _r
\log\left(\,r^3\,\lambda\,\right)\,,\label{rderiv}
\end{equation}
and
\begin{equation}
\partial _\theta\omega =-{{2\sqrt{3}\,\cot\theta}
\over{(1+\sqrt{3}\,\cot\omega\,)}}\,,\label{thetaderiv}
\end{equation}
from the integrability condition $\partial _\theta\partial
_r\omega =
\partial _r\partial _\theta\omega$ we find $\tan\omega
=-2\sqrt{3}$, implying $\cos\theta =0$, whose consistence with
(\ref{thirdeq}) is guaranteed since $\omega$ is a constant, so
that $\partial _\varphi\omega =0$. The physical meaning of this
result will be studied elsewhere.

We concentrate now on the case with $\sin\omega =0$. The
discriminant of the cubic equation (\ref{chareq}) with ${\rm tr}
K=0$, that is of
\begin{equation}
K^3 -{1\over 2}\,{\rm tr}(K^2)\,K - I \det K
=0\,,\label{simplifchareq}
\end{equation}
is found to be proportional to $\sin\omega$ as
\begin{eqnarray}
{\rm discrim}&=&\left( {{\det K}\over 2}\right) ^2 - \left(
{{{\rm tr}(K^2)}\over 6}\right) ^3\nonumber\\
&=&{{\lambda ^6}\over{64}}\,\sin ^2\,\omega\left[\,2\cos
^2\omega\,(\sin ^2\omega -4)-1\,\right]\,.\label{discrim}
\end{eqnarray}
Accordingly, condition $\sin\omega =0$ forces the solutions of
(\ref{simplifchareq}) to be three real eigenvalues summing zero,
being two of them equal to each other, as displayed in
(\ref{del11})--(\ref{del33}) below. In fact, the nontrivial
--although very simple-- dynamics for $K$ consists of equation
(\ref{firsteq}) reduced to
\begin{equation}
0= \partial _r\left(\,r^3\lambda\,\right)\,,\label{modfirsteq}
\end{equation}
while (\ref{secondeq}) and (\ref{thirdeq}) are trivially
satisfied. We solve (\ref{modfirsteq}) as
\begin{equation}
\lambda = {{2 k_0}\over{r^3}}\,,\label{lambdasol}
\end{equation}
in terms of the arbitrary constant $2 k_0$. The value
(\ref{lambdasol}) is to be replaced on the one hand in
(\ref{isotropscalconstr}), yielding
\begin{equation}
\partial _r\left[\,r\left(\,\phi -1\,\right)
+{{k_0^2}\over{r^3}} \,\right]=0\,,\label{modisotropscalconstr}
\end{equation}
giving rise, in analogy to (\ref{Schwarz2}), to the generalized
Schwarzshild solution
\begin{equation}
\phi = 1 -{{2m}\over r} -{{k_0^2}\over{r^4}}\,,\label{solution2}
\end{equation}
and on the other hand (\ref{lambdasol}), when replaced in
(\ref{paramet1})--(\ref{paramet3}) with $\sin\omega =0$, yields
the eigenvalues
\begin{eqnarray} \Delta _r{}^r &&=\hskip0.3cm\lambda
=\hskip0.2cm{{2 k_0}\over{r^3}}\,,\label{del11}\\
\Delta _\theta{}^\theta &&=-{{\lambda}\over 2}
=-{{k_0}\over{r^3}}\,,\label{del22}\\
\Delta _\varphi{}^\varphi &&=-{{\lambda}\over 2}
=-{{k_0}\over{r^3}}\,,\label{del33}
\end{eqnarray}
which determine a nontrivial extrinsic curvature not reducible to
(\ref{solution1}). Recall that this is the main result we wanted
to derive in the present paragraph. It is worth noticing that,
according to (\ref{diagonalization}) with the choice
(\ref{triadcomps}) of triads we made, we can express the
components of $K_a{}^b$ in terms of (\ref{del11})--(\ref{del33})
as
\begin{eqnarray}
K_1{}^1 &=& e_1{}^r\Delta _r{}^r e_r{}^1 =\Delta _r{}^r
=\hskip0.2cm{{2 k_0}\over{r^3}}\,,\label{k11}\\
K_2{}^2 &=& e_2{}^\theta\Delta _\theta{}^\theta e_\theta{}^2
=\Delta _\theta{}^\theta =-{{k_0}\over{r^3}}\,,\label{k22}\\
K_3{}^3 &=& e_3{}^\varphi\Delta _\varphi{}^\varphi e_\varphi{}^3
=\Delta _\varphi{}^\varphi =-{{k_0}\over{r^3}}\,,\label{k33}
\end{eqnarray}
completing the solution provided by the metric
(\ref{explicitmetric}) with (\ref{solution2}).

\section{Conclusions}

Our main result establishes that any possible solution for the
extrinsic curvature $K_{ab}$ as found out from
(\ref{scalconstrbisbis}) and (\ref{momconstrbisbis}) is defined up
to a contribution of the form (\ref{solution1}), involving the
Cotton-York tensor with the proportionality constant $k$. Thus,
the total vanishing of $K_{ab}$ requires the Cotton-York tensor to
vanish, as it is the case for general isotropic metrics. Recall
the relation existing between $K$ and the time derivative of the
three-metric \cite{Alvarez:1988tb}, namely
\begin{equation}
K_{ab}= {1\over N}\,\left(\,D_{( a}N_{b )} -{1\over 2}\,\partial
_\tau\,q_{ab}\,\right)\,,\label{timederiv}
\end{equation}
(with $N$ as the lapse function and $N_a$ as the shift vector),
showing an interplay between either vanishing or not vanishing
$K$, and the corresponding either static or time-evolving metric.

The Cotton tensor $D_{[c}\left( R_{d]}{}^b -{1\over 4}\,\delta
_{d]}^b\,R\,\right)$ present in (\ref{solution1}) is conformally
invariant under rescalings of the 3-metric
$q_{ab}\rightarrow\tilde{q}_{ab} =\varphi (x)\,q_{ab}$. Although
originally expressed as a functional of the three-metric $q_{ab}$,
(\ref{solution1}) may be alternatively formulated in different
forms, for instance as
\begin{equation}
K^I =k\,\,^\#{\buildrel\Gamma\over{\nabla}}\, \left[\,{}^\#
{\buildrel\Gamma\over {F^I}}-{1\over 2}\,\vartheta ^I\,\,{}^\#
(\,\vartheta _J\wedge {\buildrel\Gamma\over
{F^J}}\,)\,\right]\,,\label{extcalcsolution1}
\end{equation}
in terms of (\ref{FB1bis}), or equivalently as
\begin{equation}
K^I =k\,\,^\#\hat{D}\left(\,e_J\rfloor \hat{R}^{IJ} -{1\over
4}\,\vartheta ^I\,e_M\rfloor e_N\rfloor
\hat{R}^{MN}\,\right)\,,\label{extcalcsolution2}
\end{equation}
with the covariant differential built with connection
(\ref{triadconn}) --related to $\Gamma ^I$ as (\ref{PGT10})-- and
with the corresponding curvature 2-form $\hat{R}_I{}^J :=
d\,\hat{\Gamma}_I{}^J +\hat{\Gamma}_K{}^J\wedge
\hat{\Gamma}_I{}^K$ related to (\ref{FB1bis}) as
${\buildrel\Gamma\over {F^I}} = {1\over 2}\,\epsilon
^I{}_{JK}\hat{R}^{JK}$.

We have shown that the search for a general extrinsic curvature
$K_{ab}$ going beyond (\ref{solution1}) is enabled by
diagonalizing $K$ in such a way that its eigenvalues play the role
of dynamical variables. Certainly, both quantities $K_{ab}$ and
$q_{ab}$ are mutually linked to each other. However, no general
solutions $K=K(q)$ or $q=q(K)$ exist. The interplay between
$K_{ab}$ and $q_{ab}$ rather manifest itself twofold. On the one
hand, given a particular metric, restrictions result on the
eigenvalues of $K$. For instance, for the isotropic metric with
$R=0$ we found $K=0$ (Schwarzschild static solution), whereas with
$R\neq 0$ we obtained a certain $K\neq 0$, related through
(\ref{timederiv}) to the time evolution of the three-metric.
Obviously, this procedure to solve the eigenvalues of $K$ is
easily generalizable. Reciprocally, given $K$, the metric of
course is not entirely fixed by it, but the metric features are
affected by $K$ to some extent. Recall for instance
(\ref{solution2}) as the modification of (\ref{Schwarz2}) induced
by (\ref{del11})--(\ref{del33}). In general, integrability
conditions impose restrictions on the metric components, regarding
their functional form or at least their coordinate dependence.

% If in two-column mode, this environment will change to single-column
% format so that long equations can be displayed. Use
%\begin{widetext}
%\end{widetext}

\begin{acknowledgments}
The authors are very grateful to Profs. Jaime Julve, Jes\'us
Mart\'in and Friedrich W. Hehl for useful discussions and remarks.
The work of one of us (R.T.) was supported by CSIC and CESGA.
\end{acknowledgments}


\appendix

\section{\bf Nonlinear PGT connections}

Ordinary spacetime geometry, characterized by both reference
frames and connections, can be derived as a gauge theory of the
Poincar\'e group realized nonlinearly, as discussed in
\cite{Tresguerres:2002uh}. Given the linear Poincar\'e
connections, consisting of the Lorentz and translational
contributions ${\buildrel {Lor}\over{\Gamma _\alpha{}^\beta}}$ and
${\buildrel (T)\over{\Gamma ^\alpha}}$ respectively, transforming
as
\begin{equation}
\delta {\buildrel {Lor}\over{\Gamma _\alpha{}^\beta}} ={\buildrel
{Lor}\over{D}}\zeta _\alpha{}^\beta\,,\qquad \delta {\buildrel
(T)\over{\Gamma ^\alpha}} =-\zeta _\beta{}^\alpha {\buildrel
(T)\over{\Gamma ^\beta}} +{\buildrel {Lor}\over{D}}\epsilon
^\alpha\,,\label{PGT1}
\end{equation}
(where the abbreviation $Lor$ over the covariant differentials in
(\ref{PGT1}) indicates that they are constructed with the linear
Lorentz connection ${\buildrel {Lor}\over{\Gamma
^{\alpha\beta}}}$) it is characteristic for the nonlinear approach
to provide us with auxiliary Goldstone-like translational fields
$\xi^\alpha$ transforming as coordinates, namely
\begin{equation}
\delta \xi^\alpha =-\zeta _\beta {}^\alpha\,\xi^\beta -\epsilon
^\alpha\,.\label{PGT2}
\end{equation}
(Greek letters $\alpha$, $\beta$... denote internal Lorentz
indices running from $0$ to $3$.) Translational connections
modified with the help of the fields $\xi ^\alpha$ as the
nonlinear translational connections
\begin{equation}
\vartheta ^\alpha :=\,{\buildrel {Lor}\over D}\xi^\alpha
+{\buildrel (T)\over{\Gamma ^\alpha}}\,,\label{PGT3}
\end{equation}
transform as Lorentz covectors, that is
\begin{equation}
\delta \vartheta ^\alpha =-\zeta _\beta{}^\alpha \vartheta ^\beta
\,.\label{PGT4}
\end{equation}
We identify (\ref{PGT4}) as Lorentz coframe 1-forms.

On the other hand, the nonlinear approach also enables us to deal
with $SO(3)$ tensors and connections instead of with Lorentz ones,
if desired \cite{Lopez-Pinto:1997aw} \cite{rrdph}. Indeed, by
decomposing the Lorentz generators $L_{\alpha\beta}$ into space
rotations $S_I\,$ and boosts $B_I$, defined respectively as
\begin{equation}
S_I :=-\epsilon _I{}^{JK} L_{JK}\quad\,,\quad B_I :=\,2\,L_{I0}\,,
\qquad (I=\,1\,,2\,,3)\,,\label{PGT5}
\end{equation}
the components of the nonlinear Poincar\'e connection on the
corresponding Lie algebra read
\begin{equation}
{\widetilde{\Gamma}}=-i\,\hat{\vartheta}^0 P_0
-i\,\hat{\vartheta}^I P_I +i\,\Gamma ^I S_I +i\,K^I B_I
\,,\label{PGT6}
\end{equation}
where the nonlinear connections are defined with the help of boost
related Goldstone fields $\beta ^I$. Exactly as the translational
Goldstone fields $\xi ^\alpha$ are coordinate-like fields, the
boost Goldstone fields $\beta ^I$ are velocity-like fields, being
rearrangeable into the components of a Lorentz four-vector
$\left(\gamma\,,\gamma\,\beta ^I\,\right)\,$, with $\gamma
:=\,1/\sqrt{1-\beta ^2}$. In terms of them we build the
boost-transformation-analogous matrix
\begin{eqnarray}
b_0{}^0 &=&(b^{-1})_0{}^0:=\gamma\,,\nonumber\\
b_0{}^I &=&-(b^{-1})_0{}^I:=-\gamma\beta ^I\,,\nonumber\\
b_I{}^0 &=&-(b^{-1})_I{}^0:=-\gamma\beta _I\,,\nonumber\\
b_J{}^I &=&(b^{-1})_J{}^I:=\delta _J^I +(\gamma -1){{\beta _J\beta
^I}\over{\beta ^2}}\,.\label{PGT7}
\end{eqnarray}
With the help of (\ref{PGT7}) we define he nonlinear Poincar\'e
connections in (\ref{PGT6}) as
\begin{eqnarray}
\hat{\vartheta}^0&=&\,\vartheta ^\alpha b_\alpha{}^0\,, \label{PGT8}\\
\hat{\vartheta}^I&=&\,\vartheta ^\alpha b_\alpha{}^I\,, \label{PGT9}\\
\Gamma ^I&:=& {1\over 2}\,\epsilon
^I{}_{JK}\hat{\Gamma}^{JK}\nonumber\\
&=&{1\over 2}\,\epsilon ^I{}_{JK}\,(b^{-1})^{J\mu}\left(
{\buildrel {Lor}\over{\Gamma _\mu{}^\nu}}b_\nu{}^K
-d\,b_\mu{}^K\,\right)\,,\label{PGT10}\\
K^I&:=&\hat{\Gamma}^{0I}=\,(b^{-1})^{0\mu}\left( {\buildrel
{Lor}\over{\Gamma _\mu{}^\nu}}b_\nu{}^I
-d\,b_\mu{}^I\,\right)\,,\label{PGT11}
\end{eqnarray}
with $\vartheta ^\alpha$ in (\ref{PGT8}) and (\ref{PGT9}) given by
(\ref{PGT3}). Formally, (\ref{PGT8})--(\ref{PGT11}) are analogous
to gauge transformations, with the main difference that
(\ref{PGT7}) are Goldstone fields of the theory rather than group
parameters. Thus (\ref{PGT8})--(\ref{PGT11}) are definitions of
new variables whose variations are easily checked to be
\begin{eqnarray}
\delta\hat{\vartheta}^0 &=&\,0\,, \label{PGT12}\\
\delta\hat{\vartheta}^I &=&\,\epsilon ^I{}_{JK}\,{\bf{\Theta}}
^J\, \hat{\vartheta}^K\,, \label{PGT13}\\
\delta K^I &=&\,\epsilon ^I{}_{JK}\,{\bf{\Theta}} ^J\,K^K\,,\label{PGT14}\\
\delta \Gamma ^I
&=&-{\buildrel\Gamma\over{\nabla}}\,{\bf{\Theta}}^I\nonumber \\
&:=&-\left(\,d\,{\bf{\Theta}} ^I +\epsilon ^I{}_{JK}\,\Gamma
^J\,{\bf{\Theta}} ^K\,\right)\,,\label{PGT15}
\end{eqnarray}
corresponding to $SO(3)$ transformations with the group parameter
${\bf{\Theta}} ^a$ built from $\zeta _\beta {}^\alpha$ in
(\ref{PGT2}) with the help of the velocity fields as
\begin{equation}
{\bf{\Theta}} ^I:=-{1\over 2}\,\epsilon ^I{}_{JK}\left[\,\zeta
^{JK}+{{2\gamma}\over{(\gamma +1)}}\,\zeta ^{0J}\beta
^K\,\right]\,.\label{PGT16}
\end{equation}
Notwithstanding, although not explicitly displayed, the total
symmetry remains the Poincar\'e one. Notice that the tetrads
become split into an $SO(3)$ singlet $\hat{\vartheta}^0\,$ plus an
$SO(3)$ covector $\hat{\vartheta}^I$, while the Lorentz connection
decomposes into a nonlinear boost connection 1--form $K^I$
transforming as a $SO(3)$ covector plus a $SO(3)$ connections
$A^I$.

The nonlinear field strength built from (\ref{PGT6}) reads
\begin{equation}
d\,{\widetilde{\Gamma}}
+{\widetilde{\Gamma}}\wedge{\widetilde{\Gamma}}=-i\,\hat{T}^0 P_0
-i\,\hat{T}^I P_I +i\,{\cal R}^I S_I
+i\,({\buildrel\Gamma\over{\nabla}}\,K^I ) B_I\,,\label{PGT17}
\end{equation}
which plays the role of the general Poincar\'e curvature, where
the explicit form of ${\cal R}^I$ and
${\buildrel\Gamma\over{\nabla}}\,K^I$ does not differ from
(\ref{App2.23}) and (\ref{App2.22}) respectively, while the
torsion components are defined as the $SO(3)$ quantities
\begin{eqnarray}
\hat{T}^0 &:=&\,d\,\hat{\vartheta}^0
+\hat{\vartheta}_I\wedge K^I\,,\label{App2.21a}\\
\hat{T}^I &:=&\,{\buildrel\Gamma\over{\nabla}}\,\hat{\vartheta}^I
+\hat{\vartheta}^0\wedge K^I\,,\label{App2.21b}
\end{eqnarray}
with
\begin{equation}
{\buildrel\Gamma\over{\nabla}}\,\hat{\vartheta}^I
:=\,d\,\hat{\vartheta}^I +\epsilon ^I{}_{JK}\,\Gamma
^J\wedge\hat{\vartheta}^K\,.\label{App2.21c}
\end{equation}
Vanishing torsion implies that the Lorentz connection reduces to
the (anholonomic) Christoffel connection
\begin{eqnarray}
\Gamma ^{\{\}}_{\alpha\beta}&:=&\, e_{[\alpha }\rfloor
d\,\vartheta _{\beta ]} -{1\over2} \left( e_\alpha\rfloor
e_\beta\rfloor
d\,\vartheta ^\gamma\right) \vartheta _\gamma\nonumber \\
&=& -\,dx^i\, e_{[\alpha }{}^j\,\Bigl(\,\partial _i e_{j\beta
]}-\Gamma _{ij}{}^k e_{k\beta ]}\,\Bigr)\,,\label{Christoffel}
\end{eqnarray}
(with small latin letters $i$, $j$... of the middle of the
alphabet assigned to the general spacetime coordinate indices of
the underlying 4-dimensional manifold). In (\ref{Christoffel}) we
recognize the relation to the usual Riemannian language of General
Relativity involving the ordinary holonomic Christoffel symbol
\begin{equation}
\Gamma _{ij}{}^k :={1\over 2}\,g^{\,kl}\Bigl(\,\partial _i
g_{\,lj}+
\partial _j g_{\,li}-\partial _l g_{\,ij}\,\Bigr)\,,\label{Christoffel4}
\end{equation}
whose restriction to spatial slices is considered in the main body
of the present paper.

\end{document}